\documentclass[prc,nofootinbib,superscriptaddress,showpacs]{revtex4}
\usepackage{graphicx,amsmath,amssymb,bm,multirow}

\newcommand{\be}{\begin{equation}}
\newcommand{\ee}{\end{equation}}
\newcommand{\bea}{\begin{eqnarray}}
\newcommand{\eea}{\end{eqnarray}}
\newcommand{\ba}{\begin{align}}
\newcommand{\ea}{\end{align}}

\newcommand{\fmi}{\, \text{fm}^{-1}}
\newcommand{\fmiq}{\, \text{fm}^{-3}}
\newcommand{\gcmiq}{\, \text{g} \, \text{cm}^{-3}}
\newcommand{\mev}{\, \text{MeV}}
\newcommand{\vlowk}{V_{{\rm low}\,k}}

\newcommand{\oo}{\omega_1}
\newcommand{\ot}{\omega_2}
\newcommand{\qo}{{\bf q}_1}
\newcommand{\qt}{{\bf q}_2}
\newcommand{\om}{\omega}
\newcommand{\qq}{{\bf q}}
\newcommand{\kf}{k_{\rm F}}
\newcommand{\pp}{{\bf p}}
\newcommand{\kk}{{\bf k}}

\newcommand{\delp}{R_{\bf p}}

\newcommand{\xchi}{X_\sigma}
\newcommand{\wxchi}{\widetilde{X}_\sigma}

\begin{document}

\title{Unified approach to structure factors and neutrino processes
in nucleon matter}

\author{G.\ I.\ Lykasov}
\email[E-mail:~]{lykasov@jinr.ru}
\affiliation{JINR, Dubna RU-141980, Moscow Region, Russia}
\author{C.\ J.\ Pethick}
\email[E-mail:~]{pethick@nbi.dk}
\affiliation{The Niels Bohr Institute, Blegdamsvej 17, DK-2100
Copenhagen \O, Denmark}
\affiliation{NORDITA, Roslagstullsbacken 21, 10691 Stockholm, Sweden}
\author{A.~Schwenk}
\email[E-mail:~]{schwenk@triumf.ca}
\affiliation{TRIUMF, 4004 Wesbrook Mall, Vancouver, BC, V6T 2A3, Canada}


\begin{abstract}
We present a unified approach to neutrino processes in nucleon 
matter based on Landau's theory of Fermi liquids that
includes one- and two-quasiparticle-quasihole pair states 
as well as mean-field effects. 
We show how rates of neutrino processes involving two
nucleons may be calculated in terms of the collision integral
in the Landau transport equation for quasiparticles. Using a
relaxation time approximation, we solve the transport equation
for density and 
spin-density fluctuations and derive a general form for the
response functions. We apply our approach to neutral-current
processes in neutron matter, where the spin response function
is crucial for calculations of neutrino elastic and inelastic
scattering, neutrino-pair bremsstrahlung and absorption from
strongly-interacting nucleons. We calculate the relaxation rates
using modern nuclear interactions and including many-body
contributions, and find that rates of neutrino processes 
are reduced compared with estimates based on the one-pion
exchange interaction, which is used in current simulations
of core-collapse supernovae.
\end{abstract}


\maketitle

\section{Introduction}

Neutrino emission, absorption and scattering processes in nucleon 
matter play a crucial role for the physics of stellar collapse, 
supernova explosions and neutron stars~\cite{Raffelt,Prakashreview}. Since
the leptons in these processes interact weakly, the neutrino rates
can be expressed compactly in terms of the response of nuclear matter
to axial and vector probes. In many situations, the axial response
is the more important, and in this paper we concentrate on this 
case, which for a system of nonrelativistic nucleons amounts to 
the spin or spin-isospin response. These responses have been 
calculated by a number of 
groups~\cite{Sawyer,IP,Prakash1,Prakash2,BS}
allowing for single nucleon quasiparticle-quasihole pair 
states.\footnote{The basic single-particle-like excitations we
work with are quasiparticles and quasiholes that have properties
quantitatively different from those of free particles or holes.
However, for brevity, we shall refer to these excitations simply
as particles and holes.} However, this is insufficient for
rates of neutrino processes involving two nucleons, such as 
neutrino-pair bremsstrahlung and absorption, and modified Urca
reactions, in which two particle-hole pair states are necessary.
The possible importance of two particle-hole pair states for
neutrino inelastic scattering, in particular for energy exchange
and the formation of the neutrino spectra, has been emphasized by 
Raffelt {\it et al.}~\cite{Raffelt1,Raffelt2,Raffelt3}. Bounds on
the magnitude of the two particle-hole pair weight have been 
investigated in Ref.~\cite{OP} and it has been shown how the 
two-pair response is directly related to the collision term in 
Landau's transport equation for quasiparticles~\cite{LOP}.

Noncentral contributions to nuclear interactions, such as tensor
forces from pion exchanges and spin-orbit forces, are essential for
the two particle-hole pair response, as is clear from calculations
of neutrino-pair bremsstrahlung and the modified Urca processes~\cite{FM}
and from general considerations based on conservation laws~\cite{OP}.
Neutrino-pair bremsstrahlung and absorption change the number of
neutrinos and are key for equilibrating muon and tau neutrino number
densities in supernovae. The standard rates for bremsstrahlung
are based on the one-pion exchange model for nucleon-nucleon
interactions~\cite{FM} (in the context of supernovae, see for example
Ref.~\cite{Raffelt3}). This is a reasonable starting point, since
it represents the long-range part and the leading noncentral
contribution
in chiral effective field theory for nuclear forces~\cite{chiralEFT}.
However, the tensor force from pion exchange is singular at short 
distances, which in free space requires iteration in the spin-triplet 
channels~\cite{tensor}. In addition, subleading noncentral contributions
to nuclear interactions are important for reproducing nucleon-nucleon
scattering for the relevant channels and energies~\cite{Epelbaum}.

The aim of this paper is to give a unified treatment of neutrino
processes that includes one- and two-particle-hole pair states
as well as mean-field (Fermi liquid) effects consistently, and 
to present improved rate calculations of these processes
based on modern nuclear interactions beyond one-pion exchange
and including many-body contributions. A convenient framework for
doing this is Landau's theory of normal Fermi liquids. This work
represents an extension of Ref.~\cite{LOP}, which included two
particle-hole pair states only in leading order using diagrammatic
perturbation theory. Here we shall use the quasiparticle transport
equation. This provides a useful framework for understanding the
basic physics and for making detailed calculations. In this paper,
we focus on neutral-current processes in normal (nonsuperfluid) 
neutron matter. We leave for future work the application to 
mixtures of neutron and protons, charged-current reactions, and 
the extension to superfluid phases.

This paper is organized as follows. Section~\ref{nustruct} gives
an introduction to neutrino processes and the dynamical structure
factors. In Sect.~\ref{FLTtrans}, we discuss Landau Fermi-liquid
theory, show that it represents a useful effective theory for neutrino
processes in nucleon matter, and introduce the transport equation
for quasiparticles. Using a relaxation time 
approximation, we solve the transport equation for density
and spin-density 
fluctuations and derive a general form for the response 
functions in Sect.~\ref{relax}. The response function includes
contributions from one-particle-hole pair (corresponding to 
elastic scattering of neutrinos from nucleons) and 
two-particle-hole pair states (which enter calculations of 
inelastic scattering, and neutrino-pair bremsstrahlung and absorption).
In Sect.~\ref{times}, we calculate the appropriate relaxation
times for the one-pion exchange interaction and for a general
operator representation of the quasiparticle scattering amplitude.
We present results in Sect.~\ref{results} based on modern 
nuclear interactions and including many-body contributions,
and contrast these with rates obtained using the one-pion 
exchange interaction, which is typically used in 
supernova simulations. Finally, we assess the significance
of the improved treatment of nuclear interactions for 
neutrino mean free paths, energy loss and energy transfer
in supernovae. We summarize the improvements and conclude 
in Sect.~\ref{concl}.

\section{Neutrino processes and dynamical structure factors}
\label{nustruct}

For neutral-current processes, the weak interaction Lagrangian density
for low-energy probes is given by
\be
{\cal L}(x) = \frac{G_{\rm F}}{\sqrt 2} \, l_\mu(x) \, j^\mu(x) \,,
\ee
where $G_{\rm F} = 1.166 \times 10^{-5} \, {\rm GeV}^{-2}$ 
is the Fermi coupling constant and the weak neutral
currents are $l_\mu(x)$ for leptons and $j_\mu(x)$ for hadrons. The
neutrino contribution to the leptonic current is
\be
l_\mu(x) = \overline{\psi}_\nu \gamma_\mu (1-\gamma_5) \psi_\nu \,,
\ee
and for nonrelativistic nucleons the hadronic current can be written
as
\be
j_\mu(x) = \overline{\psi}_{\rm N} \gamma_\mu (C_{\rm V}-C_{\rm A}
\gamma_5) \psi_{\rm N} \approx
\phi^\dagger_{\rm N} ( C_{\rm V} \delta_{\mu 0} - C_{\rm A} 
\delta_{\mu i} \, \sigma_i ) \phi_{\rm N} \,,
\ee
where $\psi_\nu$ are neutrino fields, $\psi_{\rm N}$ nucleon
Dirac fields, $\phi_{\rm N}$ nonrelativistic nucleon spinors,
and $\sigma_i$ Pauli matrices.
The neutral-current vector coupling constant is
$C_{\rm V} = -1/2$ for neutrons and $C_{\rm V} = 1/2 - 2 \sin^2 
\theta_{\rm W} \approx 0$ for protons, $C_{\rm A}$ is the axial-vector
coupling, $C_{\rm A} = - g_a/2 = - 1.26/2$ for neutrons and
$C_{\rm A} = g_a/2$ for protons. While the vector 
current is conserved, the axial coupling can be modified in a 
many-body system. As a result, one may expect
a reduction of $g_a$ for a nucleon quasiparticle 
by $5-10 \%$ in neutron matter and 
$10-20 \%$ in symmetric nuclear matter~\cite{Arima,Cowell1}.

Consider neutrinos with incoming energy $\oo$ and momentum $\qo$
that scatter from nuclear matter to a final state with energy $\ot$
and momentum $\qt$. Since neutrinos interact weakly, the rate
for neutrino scattering
can be expressed in terms of the dynamical structure factors for
vector and axial responses of the nuclear medium~\cite{Raffelt,IP}.
Because neutron velocities in neutron matter at the densities 
of interest are nonrelativistic, these reduce to the density
and spin responses,
which are decoupled if the system is not magnetically polarized.

The dynamical structure factors depend on the energy and momentum
transferred to the system, $\om = \oo-\ot$ and $\qq = \qo-\qt$, and
are defined for the density response by~\cite{IP,Pines}
\be
S_{\rm V}(\om,\qq) = \frac{1}{\pi n} \, \frac{1}{1-e^{-\om/T}} \,
{\rm Im} \, \chi(\om,{\bf q}) 
= \frac{1}{n} \int_{-\infty}^{\infty} dt \, e^{i \om t}
\, \bigl\langle n(t,\qq) \, n(0,-\qq) \bigr\rangle \,,
\label{structdent}
\ee
and for the spin response by
\be
S_{{\rm A},ij}(\om,\qq) = \frac{1}{\pi n} \, \frac{1}{1-e^{-\om/T}} \,
{\rm Im} \, \chi_{ij}(\om,{\bf q}) 
= \frac{1}{n} \int _{-\infty}^{\infty} dt \, e^{i \om t}
\, \bigl\langle {\bf s}_i(t,\qq) \, {\bf s}_j(0,-\qq) \bigr\rangle \,,
\label{structspin}
\ee
where $n$ denotes the neutron number density, $T$ is the temperature,
${\bf s} = \phi^\dagger {\bm \sigma} \phi$ the spin density, and
$\chi(\om,\qq)$ and $\chi_{ij}(\om,\qq)$ are the density-density and
spin-density--spin-density response functions, respectively. We use 
units with $\hbar=c=k_{\rm B} = 1$.

In the long-wavelength limit, $q \to 0$, the spin response is in
the direction of the applied magnetic field, hence $\chi_{ij}=0$ for
$i \neq j$. This is not the case at non-zero $q$, and the 
transverse and longitudinal spin responses differ~\cite{IP}. However, for
neutrino processes in supernovae and neutron stars, the momentum 
transfers are small compared with typical momenta of the nucleons,
such as the Fermi momentum or the inverse Compton wavelength, and
therefore the spin response is essentially diagonal,
\be
\chi_{ij} \approx \chi_\sigma \, \delta_{ij}
\quad {\rm and} \quad
S_{{\rm A},ij} \approx S_{\rm A} \, \delta_{ij} \,.
\ee

The transition probability $\Gamma(Q_1,Q_2)$ for a neutrino with
energy and momentum $Q_1 = (\oo,\qo)$ to scatter to a state 
$Q_2 = (\ot,\qt)$ is fully 
determined by the density and spin response functions
(see for example Refs.~\cite{IP,Raffelt}),
\be
\Gamma(Q_1,Q_2) =
2\pi \, n \, G_{\rm F}^2 \, \biggl[ C_{\rm V}^2 \, (1+\cos\theta)
\, S_{\rm V}(\om,\qq) + C_{\rm A}^2 \, (3-\cos\theta) \, S_{\rm A}(\om,\qq)
\biggr] \,,
\label{cross}
\ee
where $\theta = \arccos(\widehat{\bf q}_1 
\cdot \widehat{\bf q}_2)$ is the scattering angle. The rate for bremsstrahlung
of a neutrino with four-momentum $Q_2$ and an antineutrino with four-momentum
$Q_1$ is given by $\Gamma(-Q_1,Q_2)$, and for absorption of a neutrino 
with $Q_1$ and antineutrino with $Q_2$ by $\Gamma(Q_1,-Q_2)$.
We note that Eq.~(\ref{cross}) neglects 
corrections of order $\om/m$ from weak magnetism and other
effects~\cite{Horowitz}.
In terms of the transition probability, 
the rate of change of the neutrino occupation number $n_{\qo}$
due to interaction with the nuclear medium is given by
\be
\frac{d n_{\qo}}{dt} = 
\int \frac{d\qt}{(2\pi)^3} \biggl[
\Gamma(Q_2,Q_1) n_{\qt} (1-n_{\qo})
- \Gamma(Q_1,Q_2) n_{\qo} (1-n_{\qt})
+ \Gamma(-Q_2,Q_1) (1-n_{\qo}) (1-\overline{n}_{\qt})
- \Gamma(Q_1,-Q_2) n_{\qo} \overline{n}_{\qt} \biggr] \,,
\label{dndt}
\ee
where $\overline{n}_{{\bf q}_i}$ is the antineutrino occupation number.
The four terms correspond to ``in-scattering'', ``out-scattering'', 
neutrino-pair bremsstrahlung and absorption, respectively. These differ
only by the kinematics in the dynamical structure factors.

\section{Landau Fermi-liquid theory and quasiparticle transport
equation}
\label{FLTtrans}

In supernovae and neutron stars, the neutrino energies are typically
$\omega_1, \omega_2 \lesssim 30 \mev$. The corresponding neutrino momenta 
$q_1, q_2 \lesssim 0.15 \fmi$ are therefore small compared with the 
momenta of neutrons, which are of the order of the Fermi momentum $\kf 
\sim 1.0 \fmi$ for densities $n \sim n_0/10$. Here, 
$n_0 = 0.16 \fmiq$ or $\rho_0 = 2.8 \times 10^{14} \gcmiq$ denotes
the saturation density of symmetric nuclear matter. Consequently,
it is a good first approximation to work only to lowest order in 
the neutrino momenta. In addition, we focus on situations when the
temperature is small compared with the Fermi energy of neutrons.
This is the regime in which Landau's theory of normal Fermi liquids
may be used~\cite{Landau,BaymPethick}. Landau theory provides a 
clear separation between long-wavelength, low-frequency degrees 
of freedom, which are treated explicitly, and short-wavelength,
high-frequency ones, whose effects are included in low-energy 
constants that incorporate the renormalization of matrix elements
of currents and interparticle interactions. Another strength of
Landau Fermi-liquid theory is that it brings out clearly the role played by 
conservation laws. Low-temperature expansions for Fermi liquids are
often useful for $T/\varepsilon_{\rm F} = 1/\eta \lesssim 1/\pi$. We
therefore expect our results to be reasonable for degeneracy
parameters $\eta \gtrsim 3$, which is typically valid for
the relevant densities in supernovae and neutron stars.

Nucleon matter differs from liquid $^3$He, the prototype Fermi liquid, 
in that the interactions between nucleons have significant noncentral
parts. This has several consequences. The magnetic moment of a
quasiparticle is not equal to the free space value (as discussed above,
the same holds for the axial coupling) and it is a tensor, that
depends on the orientation of the spin with respect to the momentum 
of the quasiparticle. In addition, the Landau quasiparticle interaction 
contains tensor and other noncentral contributions~\cite{noncentral},
which couple spin and orbital degrees of freedom. For the response to 
a magnetic field, which is completely equivalent to the case of an 
axial-vector probe, these effects have been explored in Ref.~\cite{OHP}.

In Landau Fermi-liquid theory, one describes the long-wavelength,
low-frequency response of the system in terms of quasiparticles.
However, if the current of interest is not a conserved
quantity, the corresponding response function at long wavelengths
contains contributions that cannot be expressed in terms of 
quasiparticle degrees of freedom. In addition, there are two-body
contributions to the effective operators.
In Ref.~\cite{OP}, it was shown
from sum-rule arguments that the contribution to the response 
not coming from single particle-hole pairs could be substantial. 
One class of processes that can be calculated within Landau Fermi-liquid
theory corresponds to creating a single particle-hole pair, which 
subsequently creates a second pair. This is taken into account
by including a collision term in the transport equation for 
quasiparticles, and in Ref.~\cite{LOP} it was described how to 
do this, starting from diagrammatic perturbation theory.  

The general formalism for calculating the rates of kinetic processes 
from microscopic theory is well developed, but to apply it to specific 
physical situations is usually complicated. However, if collisions 
are sufficiently infrequent, one can adopt an approach based on a 
kinetic equation similar to the Boltzmann equation for dilute gases, 
in which one introduces a distribution function for the elementary 
excitations that depends on the momentum of the excitation. More 
generally, when the width of an excitation becomes comparable to the
real part of the energy of an excitation, it is necessary to work in 
terms of the spectral density for adding a single particle to the 
system (the imaginary part of the single-particle propagator), which 
is a function of energy as well as of momentum~\cite{KB,Keldysh}.
In this paper, we assume that the widths are sufficiently small
that a kinetic equation approach can be used.

Next we describe the quasiparticle transport equation for a 
single-component Fermi system with spin $1/2$. We assume that
the system is not magnetically polarized. The generalization
to isospin is straightforward. The quasiparticle distribution 
function is a matrix in spin space and we write it as
\be
[n_{\bf p}]_{\alpha\alpha'} = n_{\bf p} \, \delta_{\alpha\alpha'}
+ {\bf s}_{\bf p} \cdot {\bm \sigma}_{\alpha\alpha'} \,.
\ee
Likewise, the quasiparticle energy can be written in the form
\be
[\varepsilon_{\bf p}]_{\alpha\alpha'} = \varepsilon_{\bf p} \,
\delta_{\alpha\alpha'} + {\bf h}_{\bf p} \cdot {\bm 
\sigma}_{\alpha\alpha'} \,,
\ee
where $\varepsilon_{\bf p}$ and ${\bf h}_{\bf p}$ are the 
spin-independent and spin-dependent contributions to the
quasiparticle energy. The linearized transport equation in 
momentum space for the spin response $\delta {\bf s}_{\bf p}$
of quasiparticles with momentum ${\bf p}$ is given 
by~\cite{Pines,BaymPethick}
\be
\bigl( \omega - \varepsilon_{{\bf p}+{\bf q}/2} +
\varepsilon_{{\bf p}-{\bf q}/2} \bigr) \, \delta{\bf s}_{\bf p}
+ \bigl( n_{{\bf p}+{\bf q}/2} - n_{{\bf p}-{\bf q}/2} \bigr) \,
\delta{\bf h}_{\bf p} = i \, I_\sigma[{\bf s}_{{\bf p}'}] \,,
\label{treq}
\ee
where the perturbation to the quasiparticle energy is 
\be
\delta {\bf h}_{\bf p} = {\bf U}_\sigma + 2 \int \frac{d{\bf p}'}{(2\pi)^3}
\: g_{{\bf p} {\bf p}^\prime} \, \delta {\bf s}_{\pp'} \,,
\label{qpen}
\ee
and the dependence of $\delta {\bf s}_{\pp}(\om,\qq)$ and
$\delta {\bf h}_{\pp}(\om,\qq)$ on the energy and momentum
transfers is implicit. Here, $I_\sigma[{\bf s}_{{\bf p}'}]$ is the
collision integral, the prime on the momentum argument
indicating that it generally depends on the distribution
function for states other than ${\bf p}$, 
and ${\bf U}_\sigma$ is an external field that
couples to the nucleon spin.
The spin-dependent Landau quasiparticle interaction has
a central part, $g_{{\bf p}{\bf p}'} \, {\bm \sigma}_1 \cdot 
{\bm \sigma}_2$, as well as symmetric tensor and antisymmetric
terms~\cite{noncentral}. Since the latter
are generally weaker~\cite{OHP}, we keep only the central
term in Eq.~(\ref{qpen}).
For the density response, Eq.~(\ref{treq}) 
holds with the spin-dependent contributions
replaced by their spin-independent counterparts, 
and the equation analogous to Eq.~(\ref{qpen}) is
\be
\delta \varepsilon_{\bf p} = U + 2 \int \frac{d{\bf p}'}{(2\pi)^3}
\: f_{{\bf p} {\bf p}^\prime} \, \delta n_{\pp'} \,.
\ee

In local equilibrium, the net collision rate vanishes and the
distribution function is given by the equilibrium Fermi
function for quasiparticle energy $\varepsilon_{\bf p}$,
evaluated at the values of the local chemical potential,
temperature, and flow velocity corresponding to the local
number, energy, and momentum densities. The quasiparticle
energy that occurs in the local-equilibrium distribution
function includes contributions from quasiparticle interactions,
so the quasiparticle energy is not the one for the
equilibrium state. This choice is physically the most 
meaningful, because in the energy conservation condition 
the quasiparticle energies that appear must include the 
effect of the non-equilibrium quasiparticle distribution.
From linear response theory and for $\omega=0$, the 
local-equilibrium response then follows from Eq.~(\ref{treq})
and is given by
\be
\delta {\bf s}_{\bf p} \bigr|_{\rm le} = \delp \, \delta 
{\bf h}_{\bf p} \quad {\rm with} \quad
\delp = \frac{n_{{\bf p+q}/2}-n_{{\bf p-q}/2}}{
\varepsilon_{{\bf p+ q}/2}-\varepsilon_{{\bf p-q}/2}} \,,
\ee
where the subscript ``le'' denotes the value of the quantity 
for local equilibrium.

\section{Relaxation time approximation}
\label{relax}

In general it is difficult to solve the transport equation for the full
collision integral. We therefore approximate the collision integral as
\be
I_\sigma[{\bf s}_{{\bf p}'}] =  - \frac{\delta{\bf s}_{\bf p}
- \delta{\bf s}_{\bf p} \bigr|_{\rm le}}{\tau_\sigma} \,,
\label{tau}
\ee
where $\tau_\sigma$ is an average relaxation time. 
In this section, we focus
on the spin response, but analogous expressions hold for the
density and isospin responses.
Equation~(\ref{tau}) amounts 
to the assumption that all angular harmonics of the
spin-dependent part of the quasiparticle distribution function
relax at the same rate, and this form
ensures that the collision term
vanishes when $\delta {\bf s}_{\bf p} = \delta{\bf s}_{\bf p}|_{\rm le}$.
In addition, the relaxation time is assumed to be independent of 
the quasiparticle momentum. However, consideration of the scattering
process in detail shows that, in order to obtain agreement with rates
in the collisionless limit, $|\omega| \tau_\sigma \to \infty$, the
relaxation time 
must depend on the energy transfer (see Ref.~\cite{LOP} 
and Sect.~\ref{times}).
For the spin response, $\tau_\sigma$ corresponds to the rate
of change of the nucleon spin through collisions with other nucleons,
and by solving the transport equation, we include multiple-scattering
effects.

More generally, one could have allowed for changes in the temperature 
of the two different spin components, but for Fermi systems at low
temperatures, this effect, which corresponds to thermoelectric phenomena
for charged systems, is relatively unimportant. For most condensed
matter systems, Eq.~(\ref{tau}) is a rather poor approximation,
since the total spin, which corresponds to the component of the
deviation function having angular symmetry corresponding to $l=0$
is conserved to a good approximation because noncentral forces
generally play little role, while higher-$l$ components of the
spin deviation function can decay on a much shorter timescale.
For example, in liquid $^3$He, the lack of spin conservation is 
due to the interaction between the nuclear magnetic dipole moments,
which is very weak compared with the central parts of the interatomic
interaction. However, in nuclear systems noncentral contributions
to nuclear interactions, especially those from tensor forces
due to pion exchanges, are strong and the single relaxation time
approximation is expected to be better. 
The approximate form for the collision term in the transport equation
for the density response must have a more general form, since 
particle number conservation ensures that the $l=0$ component 
of the distribution function does not relax and, for a 
single-component system, momentum conservation ensures that 
the $l=1$ component does not relax either (see for example
Ref.~\cite{HPR}). For a multi-component system, such as a mixture
of neutrons and protons, the number of particles of each
component is conserved, and consequently the $l=0$ components
cannot relax, but the $l=1$ components can relax, because
momentum may be transferred from one component to another.

\subsection{Calculation of the response function}

With the approximation Eq.~(\ref{tau}), the linearized transport 
equation can be rewritten in the following form
\be
\biggl( \omega + \frac{i}{\tau_\sigma} - {\bf v}_{\bf p} \cdot {\bf q} \biggr)
\, \delta {\bf s}_\pp + \biggl( {\bf v}_{\bf p} \cdot {\bf q} - 
\frac{i}{\tau_\sigma} \biggr) \, \delp \, \delta {\bf h}_{\bf p} = 0 \,,
\ee
with $\varepsilon_{{\bf p}+{\bf q}/2} -
\varepsilon_{{\bf p}-{\bf q}/2} \approx {\bf v}_{\bf p} \cdot {\bf q}$.
In the expansion of the quasiparticle interaction in Legendre polynomials,
the $l=0$ term $g_0$ is the dominant spin-dependent contribution in
neutron matter~\cite{RGnm}, and therefore we neglect the higher-$l$ terms.
With this assumption, the perturbation to the quasiparticle energy, 
Eq.~(\ref{qpen}), is given by
\be
\delta {\bf h}_{\bf p} = {\bf U}_\sigma + g_0 \, {\bf s}
\quad {\rm with} \quad
{\bf s} = 2 \int \frac{d{\bf p}'}{(2\pi)^3}
\: \delta {\bf s}_{\pp'} \,.
\ee
As in Eq.~(\ref{structspin}), ${\bf s}$ is the Fourier transform of 
the spin deviation. We then solve the transport equation and find
\be
{\bf s} = - \chi_\sigma(\om,\qq) \, {\bf U}_\sigma \,,
\ee
where the response function $\chi_\sigma$ is given by
\be
\chi_\sigma = \frac{\xchi}{1+g_0 \xchi} \quad {\rm and} \quad
\xchi = 2 \int \frac{d{\bf p}'}{(2\pi)^3}
\, \frac{{\bf v}_\pp \cdot {\bf q} - i/\tau_\sigma}{\omega + i/\tau_\sigma
- {\bf v}_\pp \cdot {\bf q}} \, \delp \,.
\label{chi}
\ee
Here $\xchi$ is the response function in the absence of mean-field effects.
Provided the temperature is low compared with the Fermi energy, the
main contributions to the integral in Eq.~(\ref{chi}) come from the
vicinity of the Fermi surface, which leads to
\be
\xchi = N(0) \biggl[ 1 - \frac{\om}{2 v_{\rm F} q} \ln \biggl(
\frac{\omega+ i/\tau_\sigma + v_{\rm F} q}{\omega+ i/\tau_\sigma - v_{\rm F} q}
\biggr) \biggr] \,,
\label{Xsigma}
\ee
where $N(0)=m^* \kf / \pi^2$ is the density of states at the Fermi
surface for both spin populations, $m^*$ being the nucleon effective mass
and $v_{\rm F}=\kf / m^*$ the Fermi velocity. For the imaginary
part of $\chi_\sigma$ we have
\be
{\rm Im} \chi_\sigma = \frac{{\rm Im} \, \xchi}{|1+ g_0 \xchi|^2}
= N(0) \, \frac{{\rm Im} \, \wxchi}{|1+ G_0
\wxchi|^2} \,,
\label{imchi}
\ee
with dimensionless Landau parameter $G_0 = N(0) \, g_0$, and $\wxchi
= \xchi / N(0)$, whose imaginary part is
\be
{\rm Im} \wxchi = \frac{\om}{2 v_{\rm F} q}
\biggl[ \arctan\bigl[ (\omega + v_{\rm F} q) \tau_\sigma \bigr]
- \arctan\bigl[ (\omega - v_{\rm F} q) \tau_\sigma \bigr] \biggr] \,.
\label{imchi0}
\ee
The branch of the arctangent to be used is that lying between 
$-\pi/2$ and $+\pi/2$. For $\tau_\sigma \to \infty$, the form for
${\rm Im} \chi_\sigma$ given by Eqs.~(\ref{imchi}) and~(\ref{imchi0})
reproduces the results of Ref.~\cite{IP} for single
particle-hole pair states, with
\be
{\rm Im} \wxchi \to \frac{\pi \omega}{2 v_{\rm F} q}
\: \Theta\bigl( v_{\rm F} q - |\omega| \bigr) \,,
\label{imchi0theta}
\ee
where $\Theta(x)$ is the step function. Our results
generalize earlier work by taking into account effects due to
non-zero wavelengths and recoil of the nucleons. A direct inspection
shows that the resulting dynamical structure factor satisfies the detailed
balance condition $S(-\om)=S(\om) e^{-\om/T}$. In contrast to Ref.~\cite{LOP}, 
where calculations were made to leading order in the scattering rate, 
Eq.~(\ref{imchi}) contains contributions of higher order and thereby
takes into account the Landau-Pomeranchuk-Migdal effect~\cite{LPM1,LPM2}.

In the long-wavelength limit, $q \to 0$, we have
\be
\wxchi(\om, q \to 0) = \frac{1}{1 - i \omega \tau_\sigma}
\quad {\rm and} \quad
\widetilde{\chi}_\sigma(\om, q \to 0) = \frac{1}{1+G_0 
- i \omega \tau_\sigma} \,,
\ee
with imaginary part 
\be
{\rm Im} \widetilde{\chi}_\sigma(\om, q \to 0) = \frac{\om \tau_\sigma}{
(1+G_0)^2 + (\om \tau_\sigma)^2} \,.
\label{imchiq=0}
\ee
In the absence of mean-field effects, this has the same form as the
Ansatz used by Raffelt {\it et al.} to account for multiple scattering
at low $\om$~\cite{Raffelt1,Raffelt2,Raffelt3}. 
Equation~(\ref{imchiq=0}) shows that the characteristic frequency 
for the response is $\sim (1+G_0)/\tau_\sigma$. The factor $1+G_0$ indicates
that near the transition to a ferromagnetic state, $G_0 \to -1$,
the characteristic time becomes long, corresponding to what is referred
to as critical slowing down. For neutrons, one has $G_0 > 0$~\cite{RGnm}
and the spin response is pushed to higher frequencies.

\section{Relaxation times}
\label{times}

To begin, we consider the time for an excess population of 
quasiparticles in a particular momentum, energy and spin state 
(denoted by $\pp_1$, $\varepsilon_1$ and ${\bm \sigma}_1$) to relax 
when the distribution function for all other states is that for 
equilibrium. It is convenient 
to consider the general case when the quasiparticles of the excess 
population are not on the energy shell, since this is the quantity 
which naturally enters calculations of the response functions at high
frequency~\cite{LOP}. The relaxation time can be written in operator
form 
\be
\frac{1}{\tau(\varepsilon_1+\omega,{\bm \sigma}_1 \cdot \widehat{\bf p}_1)}
=\frac{1}{\tau(\varepsilon_1+\omega)} \, (1 + \alpha \: {\bm \sigma}_1 \cdot
\widehat{\bf p}_1) \,,
\label{tauop}
\ee
where $\alpha$ is a coefficient that characterizes the strength of
noncentral contributions to the relaxation rate. Unlike in systems 
with only central interactions ($\alpha=0$), the relaxation rate depends
on the spin orientation of the quasiparticle, because spin and 
momentum are coupled.

By generalizing the standard theory of relaxation rates~\cite{BaymPethick}
to the case of noncentral interactions, we have~\cite{LOP}
\be
\frac{1}{\tau(\varepsilon_1+\omega)} = \frac{3}{4} \, C \biggl[
\, T^2 + \frac{(\varepsilon_1+\omega)^2}{\pi^2} \biggr] \,,
\label{tauC}
\ee
where the factor $3/4$ is included so that energy-averaged 
relaxation rates have a simple form (see Eqs.~(\ref{tauav}) 
and~(\ref{tauavC})) and the
coefficient $C$ is given by
\be
C = \frac{4 \pi^3}{3 N(0)^2}
\prod\limits_{i=2,3,4} \biggl( \frac{m^*}{\kf} 
\int \frac{d{\bf p}_i}{(2\pi)^3} \: \delta(p_i - \kf) \biggr) 
(2\pi)^3 \delta(\pp_1+\pp_2-\pp_3-\pp_4) \, \frac{1}{4} \, 
{\rm Tr} \bigl[ \, {\cal A}_{{\bm \sigma}_1,{\bm \sigma}_2}(\kk,\kk') \,
{\cal A}_{{\bm \sigma}_1,{\bm \sigma}_2}(-\kk,\kk') \,
\bigr] \, \biggr|_{p_1 = \kf} \:.
\label{Cdef}
\ee
Here we have taken $\pp_1$ to lie on the Fermi surface,
${\cal A}_{{\bm \sigma}_1,{\bm \sigma}_2}(\kk,\kk')$ denotes the
quasiparticle scattering amplitude in units of the density of states,
$\kk = \pp_1 - \pp_3$ and $\kk' = \pp_1 - \pp_4$ are the momentum
transfers,\footnote{We use $\kk$ and $\kk'$ for the
momentum transfers between nucleons, in order to distinguish them
from the momentum transfer $\qq$ in the structure factors. This
differs from the notation used in Refs.~\cite{noncentral,RGnm,nnbrems}
and these should also not be confused with relative momenta.}
and we have neglected the neutrino momenta in the 
delta function that expresses momentum conservation, since they 
are small compared with the Fermi momentum. The factor $1/4$ in 
Eq.~(\ref{Cdef}) is the symmetry factor.\footnote{We note that 
Refs.~\cite{FM,nnbrems} use a symmetry factor of $1/2$ instead 
of $1/4$ and consequently overestimate rates by a factor 2.} 
Since we work with antisymmetrized amplitudes one factor of $1/2$
is necessary to avoid double counting of final states, and a second
factor of $1/2$ comes from taking the average over initial spin states
of particle 1.
On the Fermi surface,
the momentum transfers are orthogonal and we can express 
Eq.~(\ref{Cdef}) as
\be
C = \frac{\pi^3 m^*}{6 \kf^2} \: \biggl\langle \: \frac{1}{4} \, 
{\rm Tr} \bigl[ \, {\cal A}_{{\bm \sigma}_1,{\bm \sigma}_2}(\kk,\kk') \,
{\cal A}_{{\bm \sigma}_1,{\bm \sigma}_2}(-\kk,\kk') \,
\bigr] \, \biggr\rangle \,,
\label{dtrace}
\ee
where the average is over the Fermi surface. In terms of $k, k'$, this
can be written as~\cite{nnbrems}
\be
\langle \, F(k,k') \, \rangle = \frac{1}{\pi}
\int\limits_0^{2 \kf} \frac{dk}{\kf} \, 
\int\limits_0^{2 \kf} \frac{dk'}{\kf} \: 
\frac{\kf \, \Theta(4 \kf^2 - k^2 - k^{\prime\,2})}{
\sqrt{4 \kf^2 - k^2 - k^{\prime\,2}}} \: F(k,k') \,.
\label{kkpav}
\ee
With this average, the coefficient $\alpha$ can be written as
\be
\alpha = \frac{1}{2} \: \frac{\biggl\langle \, 
{\rm Tr} \bigl[ \, {\bm \sigma}_1 \cdot
\widehat{\bf p}_1 \, {\cal A}_{{\bm \sigma}_1,{\bm \sigma}_2}(\kk,\kk') \,
{\cal A}_{{\bm \sigma}_1,{\bm \sigma}_2}(-\kk,\kk') \,
\bigr] \, \biggr\rangle}{\biggl\langle \,
{\rm Tr} \bigl[ \, {\cal A}_{{\bm \sigma}_1,{\bm \sigma}_2}(\kk,\kk') \,
{\cal A}_{{\bm \sigma}_1,{\bm \sigma}_2}(-\kk,\kk') \,
\bigr] \, \biggr\rangle} \,.
\ee

More general disturbances of the quasiparticle distribution function 
will depend both on the direction of the quasiparticle momentum on 
the Fermi surface and on the spin of the quasiparticle, and the 
relaxation time for the disturbance will depend on an average of 
the scattering rate over the Fermi surface and over quasiparticle 
spins, weighted by functions of the direction of the quasiparticle 
momentum and of the spin. In general, the eigenstates of the
collision operator
will have a definite value of the total angular momentum, 
which is made up of an orbital component coming from the dependence
of the quasiparticle distribution on the angle on the Fermi surface 
and of the spin of the quasiparticle.

The most important case for relaxation of long-wavelength spin 
fluctuations is a disturbance of the distribution
function corresponding to a spin polarization that is 
independent of direction on the Fermi surface. For long wavelengths 
$|\om| \gg v_{\rm F} q$ and for frequencies large compared with the
relaxation rate $|\omega| \gg 1/\tau_\sigma$, the appropriate average
relaxation time for the transport equation and the spin response is
given by~\cite{LOP}
\begin{align}
\frac{1}{\tau_\sigma} &= \frac{1}{\om N(0)} \sum\limits_{m_{s_1}}
\int \frac{d{\bf p}_1}{(2\pi)^3} \: \frac{n(\varepsilon_1) - 
n(\varepsilon_1+\om)}{\tau_\sigma(\varepsilon_1+\om,
{\bm \sigma}_1 \cdot \widehat{\bf p}_1)} \,, \\[2mm]
&= \frac{1}{\om} \int d\varepsilon_1 \,
\frac{n(\varepsilon_1) - n(\varepsilon_1+\om)}{
\tau_\sigma(\varepsilon_1+\om)} \,,
\label{tauav}
\end{align}
where the noncentral term in the spin relaxation rate
($\alpha_\sigma$ in the operator form analogous to 
Eq.~(\ref{tauop})) averages to zero. Following 
Refs.~\cite{LOP,BaymPethick}, one has for the coefficient
$C_\sigma$ for the spin relaxation rate
\be
C_\sigma = \frac{\pi^3 m^*}{6 \kf^2} \: \biggl\langle \:
\frac{1}{12} \, \sum\limits_{j=1,2,3}
{\rm Tr} \biggl[ \, {\cal A}_{{\bm \sigma}_1,{\bm \sigma}_2}(\kk,\kk') \,
{\bm \sigma}_1^j \bigl[ ({\bm \sigma}_1 + {\bm \sigma}_2)^j \, , \,
{\cal A}_{{\bm \sigma}_1,{\bm \sigma}_2}(-\kk,\kk') \bigr] \,
\biggr] \biggr\rangle \,.
\label{strace}
\ee
The commutator with the two-body spin operator demonstrates
that only noncentral terms in the scattering amplitude
contribute. The factor $1/12$ in Eq.~(\ref{strace}) includes the
symmetry factor $1/4$ and a factor $1/3$, because we have summed
over all possible directions of the spin component $j$.

Since the dependence on the quasiparticle energy factorizes from the
nuclear interaction part, we can directly calculate the average 
relaxation time of Eq.~(\ref{tauav}) and finally obtain
\be
\frac{1}{\tau} = C \, \bigl[
T^2 + (\om/2\pi)^2 \bigr]
\quad {\rm and} \quad
\frac{1}{\tau_\sigma} = C_\sigma \, \bigl[
T^2 + (\om/2\pi)^2 \bigr] \,.
\label{tauavC}
\ee

\subsection{One-pion exchange interaction}

For the one-pion exchange (OPE) 
interaction, the direct and exchange contributions
to the scattering amplitude in Born approximation are given by
\be
{\cal A}^{\rm OPE}_{{\bm \sigma}_1,{\bm \sigma}_2}(\kk,\kk')
= - N(0) \, \biggl( \frac{g_a}{2 F_\pi} \biggr)^2 \biggl[ 
\frac{{\bm \sigma}_1 \cdot \kk \, {\bm \sigma}_2 \cdot \kk}{k^2 + m^2_\pi}
- \frac{{\bm \sigma}_1 \cdot \kk' \, {\bm \sigma}_2 \cdot \kk'
+k'^2 ( 1 - {\bm \sigma}_1 \cdot {\bm \sigma}_2)/2}{k^{\prime 2} + m^2_\pi}
\biggr] \,,
\label{OPEamp}
\ee
with pion decay constant $F_\pi = 92.4 \mev$ and neutral pion mass
$m_\pi = 134.98 \mev$. The spin trace in the relaxation time for the
spin response, Eq.~(\ref{strace}), leads to
\begin{multline}
\frac{1}{12} \, \sum\limits_{j=1,2,3}
{\rm Tr} \biggl[ \, {\cal A}^{\rm OPE}_{{\bm \sigma}_1,{\bm \sigma}_2}(\kk,\kk') \,
{\bm \sigma}_1^j \bigl[ ({\bm \sigma}_1 + {\bm \sigma}_2)^j \, , \,
{\cal A}^{\rm OPE}_{{\bm \sigma}_1,{\bm \sigma}_2}(-\kk,\kk') \bigr] \,
\biggr] \\
= \frac{4}{3} \, N(0)^2 \biggl( \frac{g_a}{2 F_\pi} \biggr)^4 \biggl[ 
\frac{k^4}{(k^2 + m^2_\pi)^2} + \frac{k'^4}{(k'^2 + m^2_\pi)^2}
+ \frac{k^2 k'^2}{(k^2 + m^2_\pi) (k'^2 + m^2_\pi)} \biggr] \,.
\label{opetrace}
\end{multline}
For $m_\pi=0$, each of the three terms in the square bracket of 
Eq.~(\ref{opetrace}) yields $1$ when averaged over the Fermi
surface according to Eq.~(\ref{kkpav}), and for non-zero $m_\pi$ this
integral can be calculated analytically, and 
one finds for the spin relaxation rate from one-pion exchange~\cite{FM} 
\be
C_\sigma^{\rm OPE} = \frac{2 \pi^3 m^*}{3\kf^2} \, 
N(0)^2 \biggl( \frac{g_a}{2 F_\pi} \biggr)^4 \, G\biggl(\frac{m_\pi}{
2\kf}\bigg) \,,
\label{tausope}
\ee
where the factor $G(x)$ takes into account the effects of a 
non-zero pion mass,
\be
G(x) = 1 -\frac{5x}{3} \, \arctan \biggl(\frac{1}{x}\biggr) +
\frac{x^2}{3(1+x^2)} + \frac{x^2}{3 \sqrt{1+2x^2}} \, 
\arctan \biggl(\frac{\sqrt{1+2x^2}}{x^2}\biggr) \,.
\ee
For $|\om| \tau_\sigma \gg 1$, the imaginary part of the spin response
function in the long-wavelength limit is given by $N(0)/(\om 
\tau_\sigma)$ (see Eq~(\ref{imchiq=0})).
In this limit, when multiple-scattering effects are small, 
our result for the dynamical
structure factor using the spin relaxation time of Eq.~(\ref{tausope})
agrees with the result of Raffelt {\it et al.}~\cite{Raffelt1,Raffelt2}
using $f/m_\pi \approx g_a/2 F_\pi$.

We can compare the spin relaxation time $\tau_\sigma^{\rm OPE}$ with
the relaxation time corresponding to decay of an excess of 
quasiparticles in a particular momentum state $\tau^{\rm OPE}$.
For the latter, the spin trace of Eq.~(\ref{opetrace}) 
has to be replaced by the one in the brackets $\langle \ldots \rangle$ 
of Eq.~(\ref{dtrace}), which yields exactly
the same result as the right-hand side of Eq.~(\ref{opetrace}) up
to the factor $4/3$. As a result, we find that the spin relaxation
rate and thus spin-flipping collisions obtained from the one-pion
exchange interaction are comparable to the relaxation rate for 
decay of an excess population in one momentum state, with
\be
\frac{1}{\tau_\sigma^{\rm OPE}} = \frac{4}{3} \, 
\frac{1}{\tau^{\rm OPE}} \,.
\label{OPEratio}
\ee
This result highlights the importance of noncentral contributions to
nuclear interactions and encourages us to perform more systematic
calculations of these rates beyond one-pion exchange. Next, we calculate
the contributions to the relaxation times from a general representation
of the quasiparticle scattering amplitude and
present results in Sect.~\ref{results}.

\subsection{General operator representation}

For neutron matter, using the general operator representation of the
scattering amplitude on the Fermi surface in the notation of 
Refs.~\cite{noncentral,nnbrems}, we find for the spin trace of
Eq.~(\ref{strace}):\footnote{We note that the factor $3$ in front of
the cross vector amplitude in Eq.~(7) of Ref.~\cite{nnbrems} should
be $1$.}
\begin{multline}
\frac{1}{12} \, \sum\limits_{j=1,2,3}
{\rm Tr} \biggl[ \, {\cal A}_{{\bm \sigma}_1,{\bm \sigma}_2} \,
{\bm \sigma}_1^j \, \bigl[ ({\bm \sigma}_1 + {\bm \sigma}_2)^j \, , \,
{\cal A}_{{\bm \sigma}_1,{\bm \sigma}_2} \bigr] \,
\biggr] \\[1mm]
= \frac{4}{3} \, 
\bigl[ \, \widetilde{\cal A}^{\,2}_{\text{tensor}}
+ \widetilde{\cal A}^{\,2}_{\text{exch. tensor}}
- \widetilde{\cal A}_{\text{tensor}}
\, \widetilde{\cal A}_{\text{exch. tensor}}
+ {\cal A}^{\,2}_{\text{spin-orbit}}
+ {\cal A}^{\,2}_{\text{diff. vector}}
+ {\cal A}^{\,2}_{\text{cross vector}} \, \bigr] \,,
\label{genstrace}
\end{multline}
where the amplitudes on the right-hand side are functions of
$k$ and $k'$. The scattering amplitudes on the Fermi surface
$\widetilde{\cal A}_{\text{tensor}}$,
$\widetilde{\cal A}_{\text{exch. tensor}}$, ${\cal A}_{\text{spin-orbit}}$,
${\cal A}_{\text{diff. vector}}$ and ${\cal A}_{\text{cross vector}}$
are real and characterize 
the momentum-dependent strengths (in units of the density of states)
of the tensor operator
$S_{12}(\widehat{\kk})$, the exchange tensor $S_{12}(\widehat{\kk}')$,
the spin-orbit operator $i ({\bm \sigma}_1 + {\bm \sigma}_2) 
\cdot \widehat{\kk} \times \widehat{\kk'}$,
the spin difference vector $i ({\bm \sigma}_1 - {\bm \sigma}_2) 
\cdot \widehat{\kk} \times \widehat{\bf P}$ (or antisymmetric spin-orbit),
and the cross vector operator $({\bm \sigma}_1 \times {\bm \sigma}_2) 
\cdot (\widehat{\kk'} \times \widehat{\bf P})$, respectively, with 
two-body center-of-mass momentum ${\bf P} = \pp_1 + \pp_2
= \pp_3 + \pp_4$ (for details, see Refs.~\cite{noncentral,nnbrems}).
The latter two operators do not conserve the spin of the interacting
particle pair and are induced in the medium due to screening by 
particle-hole excitations~\cite{noncentral}. Finally, the tilde on
the tensor parts of the scattering amplitude indicates that they
take into account induced center-of-mass tensor operator 
contributions, since this is not a linearly-independent operator 
on the Fermi surface, as discussed in Refs.~\cite{noncentral,nnbrems}.

For the spin trace of Eq.~(\ref{dtrace}) corresponding to the 
relaxation rate for decay of an excess population in one momentum 
state, we have
\begin{align}
\frac{1}{4} \, 
{\rm Tr} \bigl[ \, {\cal A}_{{\bm \sigma}_1,{\bm \sigma}_2} \,
{\cal A}_{{\bm \sigma}_1,{\bm \sigma}_2} \, \bigr]
&= {\cal A}^{\,2}_{\text{scalar}} + 3 \, {\cal A}^{\,2}_{\text{spin}}
+ \frac{2}{3} \, \biggl[ \, \widetilde{\cal A}^{\,2}_{\text{tensor}}
+ \widetilde{\cal A}^{\,2}_{\text{exch. tensor}}
- \widetilde{\cal A}_{\text{tensor}}
\, \widetilde{\cal A}_{\text{exch. tensor}} \, \biggr] \nonumber \\[2mm]
&+ 2\, {\cal A}^{\,2}_{\text{spin-orbit}}
+ 2 \, {\cal A}^{\,2}_{\text{diff. vector}}
+ 2\, {\cal A}^{\,2}_{\text{cross vector}} \,,
\label{gendtrace}
\end{align}
where in addition the central parts of the scattering 
amplitude, ${\cal A}_{\text{scalar}}$ and ${\cal A}_{\text{spin}}$,
contribute. These correspond to the spin-independent amplitude
and the spin-spin operator ${\bm \sigma}_1 \cdot {\bm \sigma}_2$,
respectively. We note that all contributions in Eqs.~(\ref{genstrace})
and~(\ref{gendtrace}) are positive. The minus sign of the direct-exchange
tensor interference term is canceled by a relative minus sign in the exchange
tensor amplitude.

\section{Results}
\label{results}

We calculate the contributions beyond one-pion exchange based
on low-momentum interactions $\vlowk$~\cite{Vlowk,smooth}, which
are obtained by evolving nuclear forces to low momentum using
the renormalization group. The resulting two-nucleon
interactions become universal at momentum scales $\Lambda 
\lesssim 2 \fmi$ and provide a basis for model-independent 
predictions of low-energy processes. The renormalization-group
evolution preserves the long-range parts from pion exchanges
and $\vlowk$ includes subleading noncentral contributions, so that all
low-energy nucleon-nucleon scattering observables and deuteron
properties are reproduced. In this first study, we have not 
included contributions from low-momentum three-nucleon 
interactions~\cite{Vlowk3N}. Their effects are generally weaker 
in neutron matter, but calculations of the equation of state show that 
three-nucleon interactions become important for $\kf \gtrsim 1.5 
\fmi$~\cite{neutmatt}. We will study their contributions to
neutrino processes in future work.

In addition, we include many-body noncentral and central
correlations from second-order particle-particle (plus hole-hole)
and particle-hole contributions using the same $\vlowk$
interactions. The resulting
quasiparticle scattering amplitudes are discussed in detail in 
Ref.~\cite{noncentral} and have been used to calculate the
neutrino emissivity from pair bremsstrahlung for neutron star
cooling~\cite{nnbrems}. Based on our results and general
arguments~\cite{nucmatt}, second-order corrections become
reasonable for low-momentum interactions. The 
intermediate states include all possible excitations for 
interacting particles on the Fermi surface. We use the
effective mass obtained from the lowest-order $\vlowk$
for all results, including for the estimates based on the
one-pion exchange interaction. The effective mass 
varies from $m^*/m = 0.95$ 
at $\kf = 1.0 \fmi$ to $m^*/m = 0.78$ at $\kf = 2.0 \fmi$, and 
in this range is well approximated by a linear dependence
on the Fermi momentum. We note that one expects an increase 
of the effective mass due to polarization effects, but this
is compensated by the reduction of the quasiparticle strength $z_{\kf}$,
as can be seen from the results of the renormalization-group
calculation of induced interactions in neutron matter~\cite{RGnm}.
We emphasize that a second-order calculation cannot give 
final results, but it provides a range for the effects due
to many-body correlations.

Finally, we note that the effect of particle-particle correlations
on neutrino-pair bremsstrahlung and other neutrino processes has
been investigated previously in Refs.~\cite{Blaschke,Hanhart,Dalen,Cowell2}.

\subsection{Relaxation times}
\label{res:times}

\begin{figure}[t]
\begin{center}
\includegraphics[scale=0.4,clip=]{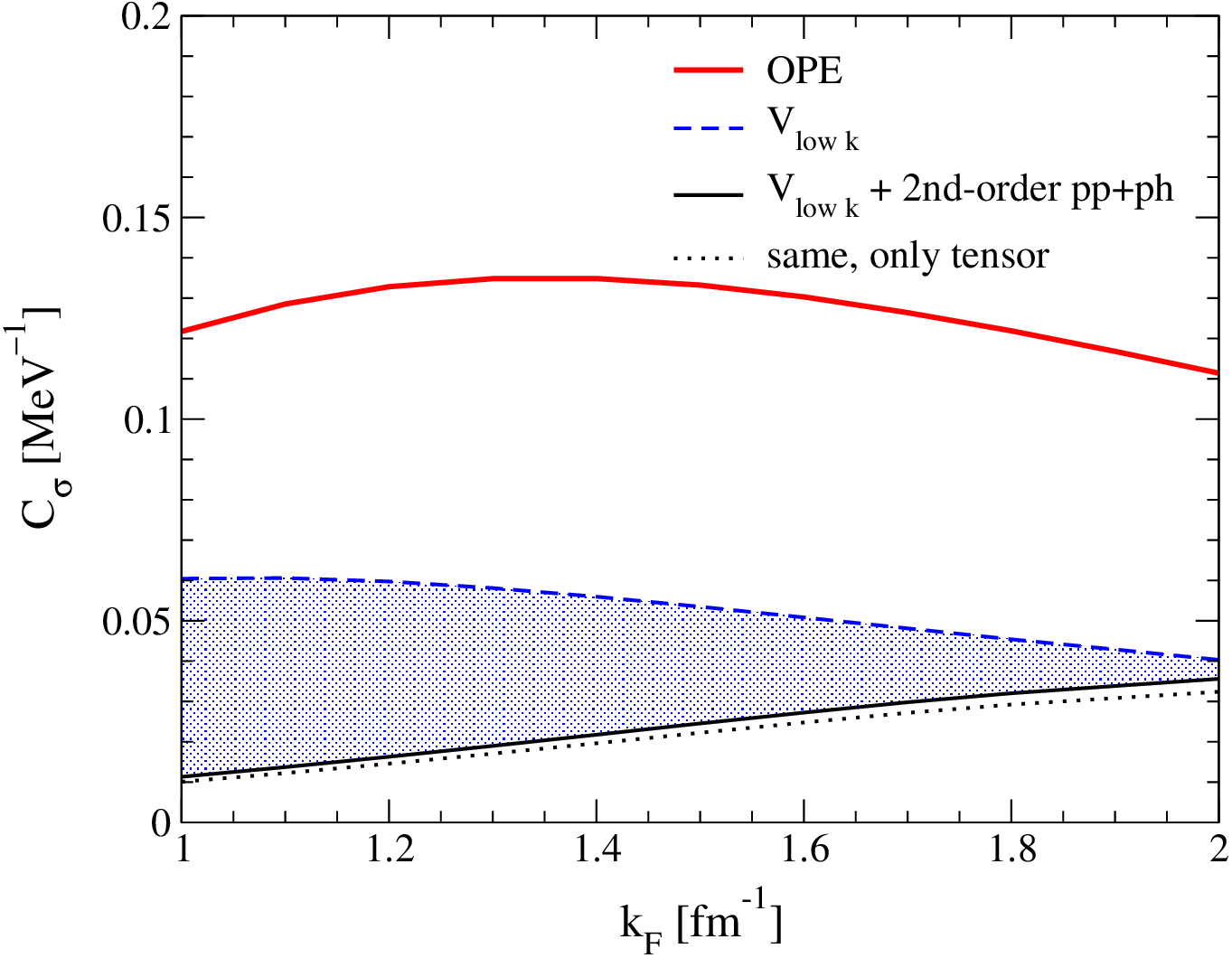}
\end{center}
\caption{(Color online)
The spin relaxation rate given by $C_\sigma$
of Eq.~(\ref{strace}) as a function of Fermi momentum $\kf$ obtained
from the one-pion exchange interaction (OPE), from low-momentum 
interactions $\vlowk$, and including second-order many-body contributions.
In addition, we show that the result obtained from $\vlowk$ plus 
second-order contributions is dominated by tensor interactions
(dotted versus solid line).\label{fig:spin}}
\end{figure}

Our results for the spin relaxation coefficient $C_\sigma$ of 
Eq.~(\ref{strace}) are shown in Fig.~\ref{fig:spin}. For energies $\om = 0$
and $T=5-10 \mev$, the value of $C_\sigma = 0.1 \mev^{-1}$ corresponds
to spin relaxation rates $1/\tau_\sigma = 2.5-10 \mev$. We find 
that the OPE model significantly overestimates the strength of noncentral 
contributions, compared to low-momentum interactions $\vlowk$, for
all considered densities. Beyond the $\vlowk$ results, we find that
second-order many-body contributions reduce the spin relaxation
rate especially at lower densities (note that $C_\sigma$ is
proportional to the square of the quasiparticle scattering amplitude).
These effects are due to second-order particle-hole interference
of tensor with strong central interactions, which are driven by large 
scattering lengths at very low densities. The band in Fig.~\ref{fig:spin}
from $\vlowk$ to including second-order contributions provides a 
range for the effects due to many-body correlations. In addition,
we observe that the spin relaxation rate depends only weakly on
density, and the rate obtained from $\vlowk$ plus second-order 
contributions is dominated by the tensor terms in Eq.~(\ref{genstrace}).

\begin{figure}[t]
\begin{center}
\includegraphics[scale=0.4,clip=]{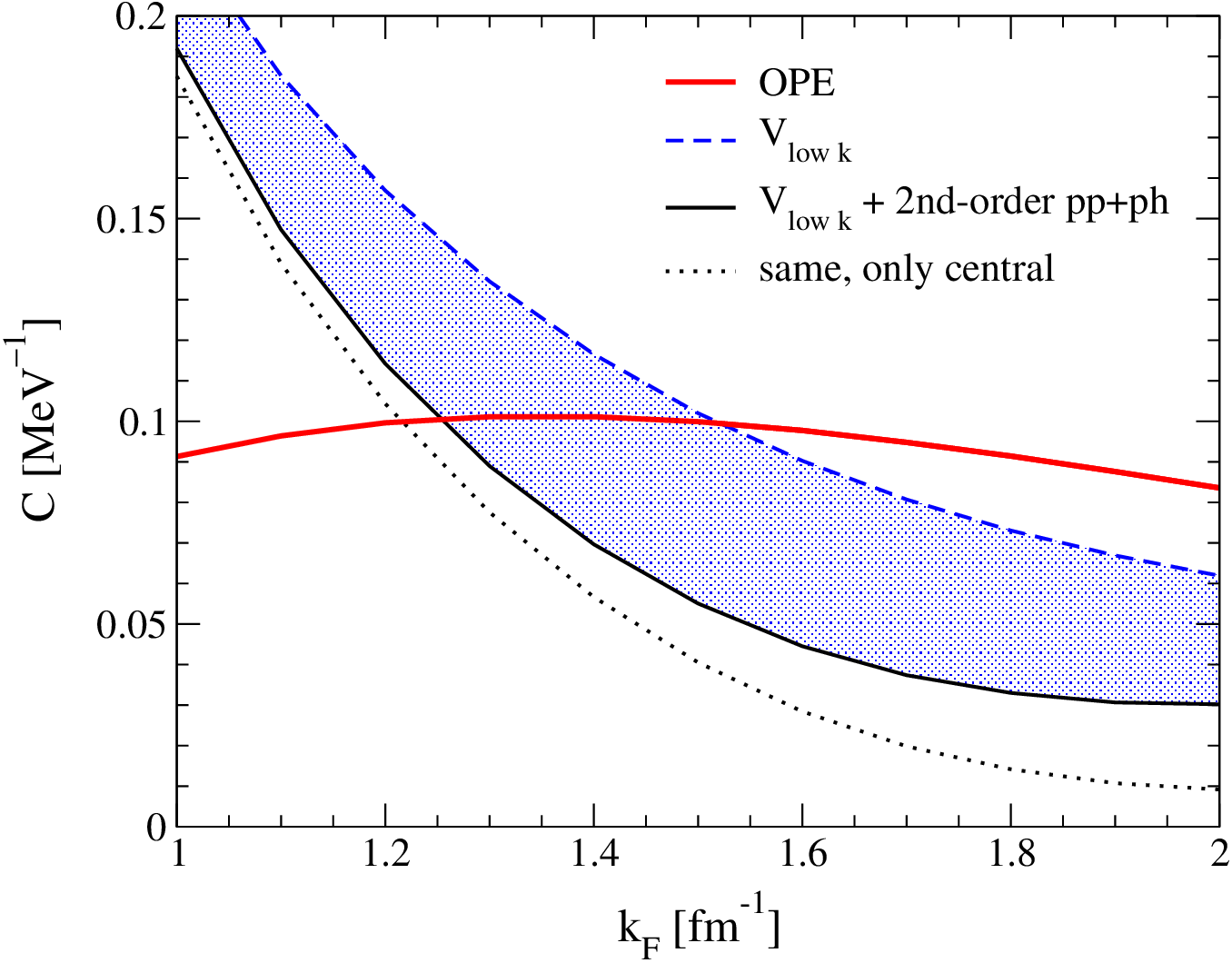}
\end{center}
\caption{(Color online)
The relaxation rate for decay of an excess of quasiparticles in a 
particular momentum state given by 
$C$ of Eq.~(\ref{dtrace}) as a function of Fermi momentum $\kf$ obtained
from the one-pion exchange interaction (OPE), from low-momentum 
interactions $\vlowk$, and including second-order many-body contributions.
In addition, we show that the result obtained from $\vlowk$ plus 
second-order contributions is dominated by central interactions
(dotted versus solid line).\label{fig:dens}}
\end{figure}

For the relaxation coefficient $C$ of Eq.~(\ref{dtrace}) corresponding 
to decay of an excess of quasiparticles in a particular momentum state, 
we obtain rates in Fig.~\ref{fig:dens} that 
are of similar magnitude compared with the spin relaxation rate.
While the OPE rate is approximately independent of density, the OPE
model underestimates the relaxation rate at low densities. This is
because the central part of the OPE interaction
$\sim k^2$ and $\sim k'^2$ 
does not capture the central shorter-range physics in nuclear forces.
This deficiency of the OPE model is most prominent at low densities, 
in comparison
to the increasing $\vlowk$ rate. Similar to the spin response, we find
a reduction of $C$ due to second-order many-body contributions, where
the band in Fig.~\ref{fig:dens} again indicates a range for the effects
due to many-body correlations. Finally, as expected, the 
relaxation rate obtained from $\vlowk$ plus second-order 
contributions is now dominated by the central terms in Eq.~(\ref{gendtrace}).

\begin{figure}[t]
\begin{center}
\includegraphics[scale=0.4,clip=]{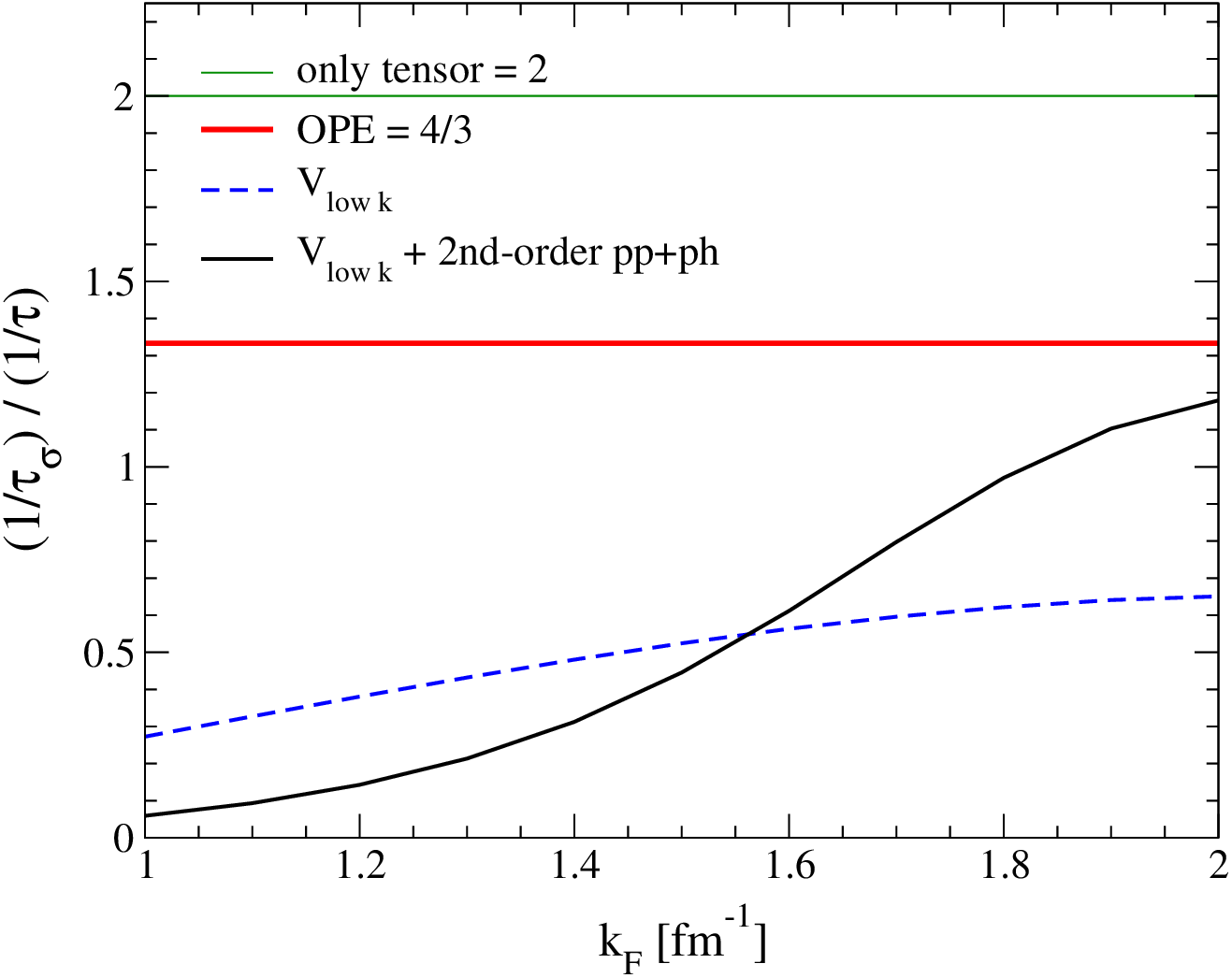}
\end{center}
\caption{(Color online)
Ratio of the spin relaxation rate to the relaxation rate for an excess of 
quasiparticles in a single momentum state
$(1/\tau_\sigma)/(1/\tau)$ as a function of Fermi momentum $\kf$ for
purely tensor scattering amplitudes (in which case the value is 2), 
for the one-pion exchange interaction
(which gives the value $4/3$), from low-momentum 
interactions $\vlowk$, and including second-order many-body contributions.
\label{fig:ratio}}
\end{figure}

\begin{figure}[t]
\begin{center}
\includegraphics[scale=0.4,clip=]{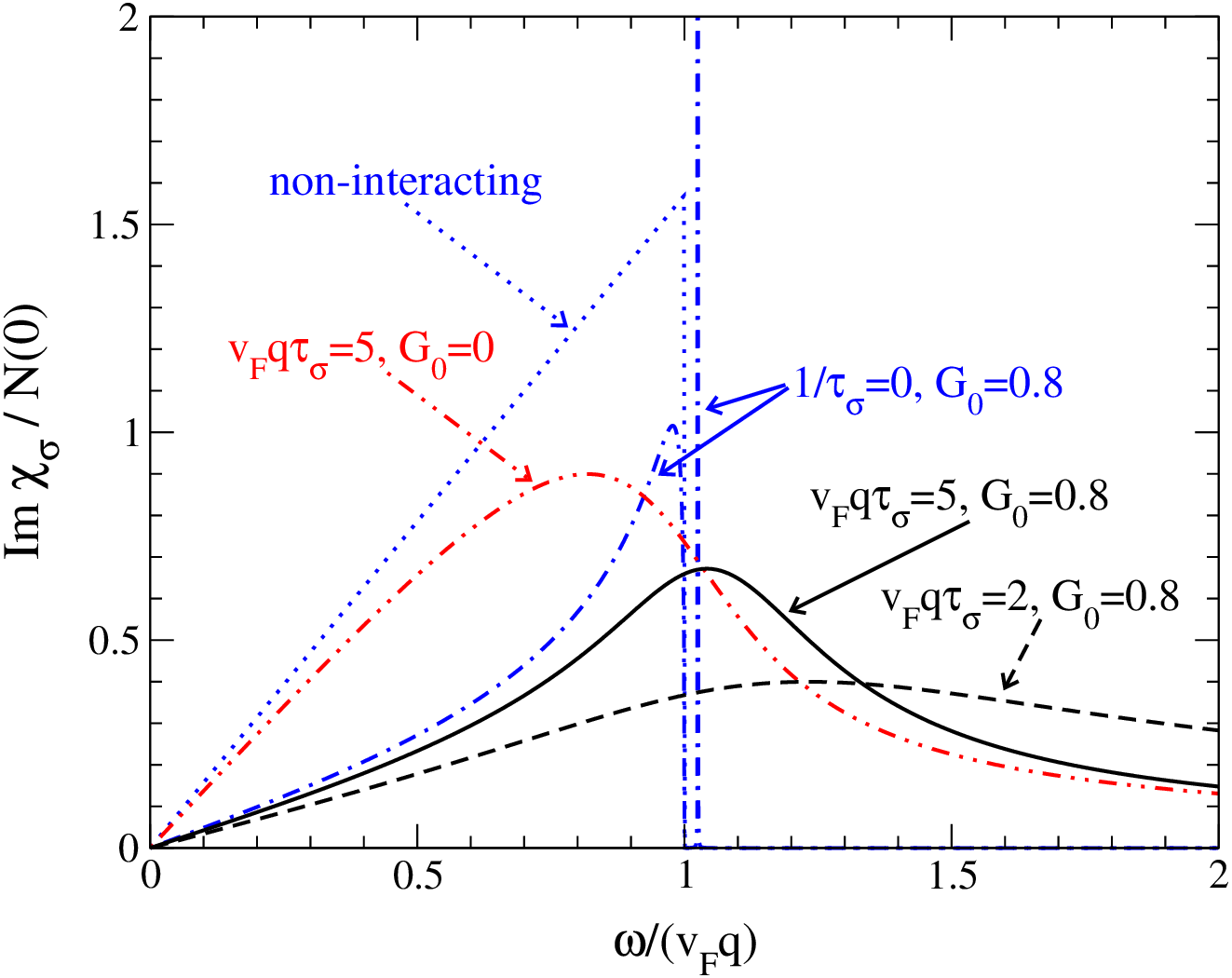}
\end{center}
\caption{The imaginary part of the spin response function ${\rm Im} 
\chi_\sigma/N(0)$ of Eq.~(\ref{imchi}) in units of the density
of states versus $\om/(v_{\rm F} q)$. Results are shown for
the non-interacting system, without and with mean-field effects, $G_0
= 0$ and $G_0 = 0.8$ respectively, and for different values of the spin
relaxation rate $1/\tau_\sigma = 0$, $v_{\rm F} q \tau_\sigma = 2$
and $v_{\rm F} q \tau_\sigma = 5$.\label{fig:sfact}}
\end{figure}

In Fig.~\ref{fig:ratio} we show the ratio $(1/\tau_\sigma)/(1/\tau)$
of the spin relaxation rate to the relaxation rate for an excess of 
quasiparticles in a single momentum state as a function
of Fermi momentum $\kf$. This is a very useful measure of the strength of
noncentral interactions compared to central ones. For purely tensor 
scattering amplitudes,
the ratio of the corresponding spin traces in Eqs.~(\ref{genstrace}) 
and~(\ref{gendtrace}) gives $(1/\tau_\sigma)/(1/\tau) = 2$, while for
the OPE interaction, which has a central part in Eq.~(\ref{OPEamp}),
this ratio is $(1/\tau_\sigma)/(1/\tau) = 4/3$, see Eq.~(\ref{OPEratio}).
While the ratio obtained
from $\vlowk$ and including second-order many-body contributions is 
considerably smaller at low densities, the relative strength of noncentral 
interactions increases with momentum and thus with density, as can be 
seen in the results of Fig.~\ref{fig:ratio} based on modern nuclear 
forces.

\subsection{Dynamical structure factor}

Motivated by the importance for neutrino rates, 
we focus on the spin response
in this section. The dynamical structure factor is determined by the
imaginary part of the spin response function ${\rm Im} \chi_\sigma$,
which is given by Eq.~(\ref{imchi}) in the relaxation time 
approximation. In units of the density of states, the
imaginary part  ${\rm Im} \widetilde{\chi}_\sigma$ is a function
of $v_{\rm F} q \tau_\sigma$ and $\omega/(v_{\rm F} q)$ or of 
$v_{\rm F} q \tau_\sigma$ and $\omega \tau_\sigma$. In the 
long-wavelength limit, $q \to 0$, we have already found that this 
is proportional to $\omega$ times a Lorentzian  function of $\om$, 
see Eq.~(\ref{imchiq=0}). Therefore, we plot in Fig.~\ref{fig:sfact}
the imaginary part of the spin response function versus 
$\om/(v_{\rm F} q)$. Results are shown for the non-interacting
system, without and with mean-field effects, $G_0 = 0$ and $G_0 
= 0.8$ respectively, and for different values of the spin
relaxation rate $1/\tau_\sigma = 0, v_{\rm F} q/5,$ and $v_{\rm F} q/2$.
We have taken the Landau
parameter from renormalization-group calculations of induced 
interactions in neutron matter~\cite{RGnm}, which yield $G_0
\approx 0.8$ over the densities considered in Sect.~\ref{res:times}.
The values of $v_{\rm F} q \tau_\sigma=2-5$ correspond to spin 
relaxation rates based on Fig.~\ref{fig:spin} for typical 
momentum transfers $q \sim \omega$ over the range $T=5-
10 \mev$ and $\kf=1.0-1.7 \fmi$. With $1/\tau_\sigma$
comparable to $v_{\rm F} q$, these estimates also show
that recoil effects may be important.

In the non-interacting case, $G_0=0$ and $1/\tau_\sigma = 0$,
the imaginary part of the spin response function is given by
$\pi \om/(2 v_{\rm F} q)$ times a step function, see
Eq.~(\ref{imchi0theta}). With single-pair mean-field effects,
$G_0 = 0.8$, a collective spin-zero-sound mode appears
as a pole contribution at $\omega/(v_{\rm F} q)|_{\rm zs} > 1$,
where the position of the pole is given by~\cite{IP}
\be
1+ G_0 \, \wxchi( \omega/(v_{\rm F} q)|_{\rm zs} \, , 
1/\tau_\sigma =0 ) = 0 \,.
\ee
As the spin relaxation rate increases, going from $1/\tau_\sigma = 0$
to $v_{\rm F} q /5$ and $v_{\rm F} q /2$, the
response is pushed to higher frequencies and the spin-zero-sound peak
disappears already for these moderate spin relaxation rates. 
For comparison, we also
show the effects due to single-pair states at $v_{\rm F} q \tau_\sigma 
= 5$, where interactions ($G_0 = 0.8$) decrease the response
at low $\om/(v_{\rm F} q)$ and also move the strength to higher 
frequencies.

\subsection{Neutrino mean free paths, energy loss and energy transfer}

We next assess the significance of the improved rates for neutrino 
mean free paths, energy loss and energy transfer. For derivations of
Eqs.~(\ref{imfp}) to~(\ref{tscatt}) see Refs.~\cite{Raffelt,Raffelt3}.
All rates are for one neutrino flavor.
We emphasize that the OPE results are based on the solution to the
transport equation in the relaxation time approximation, and do not
correspond directly to OPE rates used in supernova simulations.
For simple
estimates, we use the dynamical structure factor for spin fluctuations
in the long-wavelength limit, $S_\sigma(\omega) = S_\sigma(\omega,
q \to 0)$, given by Eqs.~(\ref{structspin}) and~(\ref{imchiq=0}),
without further approximations or Ansaetze for the structure factor.
Effects due to the finite wavelength and recoil of the nucleons will
be studied in future work.

\begin{table}[t]
\begin{tabular}{ll|cc|cc|cc}
\hline\hline
& $G_0$ & $\quad\quad 0 \quad\quad$ & $\quad\quad 0.8 \quad\quad$ 
& $\quad\quad 0 \quad\quad$ & $\quad\quad 0.8 \quad\quad$ 
& $\quad\quad 0 \quad\quad$ & $\quad\quad 0.8 \quad\quad$ \\ \hline
$\kf \; [{\rm fm}^{-1}]\quad$ & $T \; [{\rm MeV}]\quad$ &
\multicolumn{2}{c|}{$C_\sigma$ from OPE} &
\multicolumn{2}{c|}{$\vlowk$} &
\multicolumn{2}{c}{$\vlowk$ + 2nd order} \\ \hline\hline
\multirow{2}{*}{1.0} & 5 & 0.0770 & 0.0697 & 0.0397 & 0.0386 & 
0.00754 & 0.00753 \\
& 10 & 1.08 & 0.798 & 0.612 & 0.554 & 0.120 & 0.120 \\ \hline
\multirow{2}{*}{1.7} & 5 & 0.119 & 0.107 & 0.0476 & 0.0468 & 0.0296 & 0.0294 \\
& 10 & 1.66 & 1.21 & 0.744 & 0.700 & 0.470 & 0.457 \\ \hline\hline
\end{tabular}
\caption{Thermally averaged inverse neutrino 
mean free path $\langle \lambda^{-1}
\rangle$ in ${\rm km}^{-1}$ calculated from 
Eq.~(\ref{imfp}) for characteristic temperatures and Fermi momenta.
Results are given without and with mean-field effects, $G_0 = 0$ and 
$G_0 = 0.8$ respectively, and for different spin relaxation rates 
$1/\tau_\sigma$ based on Fig.~\ref{fig:spin}.\label{tab:imfp}}
\end{table}

In Table~\ref{tab:imfp} we present results for an average
inverse neutrino mean free path $\langle \lambda^{-1} \rangle$,
\be
\langle \lambda^{-1} \rangle = \frac{C_{\rm A}^2
G_{\rm F}^2}{20 \pi} \, \frac{n}{T^3} \int\limits_0^\infty d\omega
\, \omega^5 \, e^{-\omega/T} \, S_\sigma(\omega) \,,
\label{imfp}
\ee
for characteristic temperatures and Fermi momenta. This result 
applies for a Maxwellian initial distribution of neutrinos, and 
Pauli blocking in the final state has been ignored. We consider
structure factors without and with mean-field effects, 
$G_0 = 0$ and $G_0 = 0.8$ respectively, and for different 
spin relaxation rates $1/\tau_\sigma$ based on Fig.~\ref{fig:spin}. 
With the spin relaxation rates obtained from $\vlowk$ and including
second-order many-body contributions, the mean free paths are
significantly longer compared to the OPE model. This follows the
reduction of $C_\sigma$ seen in Fig.~\ref{fig:spin}. For OPE, 
the effects of interactions ($G_0=0.8$ compared to $G_0=0$) 
reduce the neutrino scattering rate, especially at higher
temperature. In contrast, with the rates based on 
low-momentum interactions, $\omega \tau_\sigma$ is larger
and the imaginary part of the spin response function approaches
${\rm Im} \chi_\sigma(\om, q \to 0) \to N(0)/(\om \tau_\sigma)$.
As a result, mean-field effects are weak for $|\omega| \tau_\sigma
\gg 1$ in the long-wavelength limit.

\begin{table}[t]
\begin{tabular}{ll|cc|cc|cc}
\hline\hline
& $G_0$ & $\quad\quad 0 \quad\quad$ & $\quad\quad 0.8 \quad\quad$ 
& $\quad\quad 0 \quad\quad$ & $\quad\quad 0.8 \quad\quad$ 
& $\quad\quad 0 \quad\quad$ & $\quad\quad 0.8 \quad\quad$ \\ \hline
$\kf \; [{\rm fm}^{-1}]\quad$ & $T \; [{\rm MeV}]\quad$ &
\multicolumn{2}{c|}{$C_\sigma$ from OPE} &
\multicolumn{2}{c|}{$\vlowk$} &
\multicolumn{2}{c}{$\vlowk$ + 2nd order} \\ \hline\hline
\multirow{2}{*}{1.0} & 5 & 1.77 & 1.62 & 0.911 & 0.888 & 0.173 & 0.172 \\
& 10 & 4.02 & 3.00 & 2.25 & 2.06 & 0.441 & 0.440 \\ \hline
\multirow{2}{*}{1.7} & 5 & 2.75 & 2.49 & 1.09 & 1.07 & 0.679 & 0.675 \\
& 10 & 6.18 & 4.55 & 2.73 & 2.57 & 1.72 & 1.68 \\ \hline\hline
\end{tabular}
\caption{Energy-loss rate $Q$ of Eq.~(\ref{eloss}) due to neutrino-pair
bremsstrahlung, $nn \to 
nn \nu \overline{\nu}$, for characteristic temperatures and Fermi momenta.
Results are given without and with mean-field effects, $G_0 = 0$ and 
$G_0 = 0.8$ respectively, and for different spin relaxation rates 
$1/\tau_\sigma$ based on Fig.~\ref{fig:spin}. The energy-loss rates
are in units of $10^{33} \, {\rm erg} \, {\rm cm}^{-3} \, {\rm s}^{-1}$
for $T=5 \mev$ and  $10^{35} \, {\rm erg} \, {\rm cm}^{-3} \, 
{\rm s}^{-1}$ for $T=10 \mev$.\label{tab:eloss}}
\end{table}

The energy-loss rate $Q$ due to neutrino-pair bremsstrahlung, $nn \to 
nn \nu \overline{\nu}$, of neutron matter transparent to neutrinos
is given by
\be
Q = \frac{C_{\rm A}^2 G_{\rm F}^2 \, n}{20 \pi^3} \, 
\int\limits_0^\infty d\omega
\, \omega^6 \, e^{-\omega/T} \, S_\sigma(\omega) \,.
\label{eloss}
\ee
Our results for the energy-loss rate $Q$ are listed in Table~\ref{tab:eloss}
for characteristic temperatures, Fermi momenta and the different
cases of the structure factor. They follow the same general pattern
as the inverse mean free paths in Table~\ref{tab:imfp}: 
a reduction of the energy loss calculated with modern nuclear forces compared
to OPE and consequently weak mean-field effects.

\begin{table}[t]
\begin{tabular}{ll|cc|cc|cc|c}
\hline\hline
& $G_0$ & $\quad\quad 0 \quad\quad$ & $\quad\quad 0.8 \quad\quad$ 
& $\quad\quad 0 \quad\quad$ & $\quad\quad 0.8 \quad\quad$ 
& $\quad\quad 0 \quad\quad$ & $\quad\quad 0.8 \quad\quad$ & \\ \hline
$\kf \; [{\rm fm}^{-1}]\quad$ & $T \; [{\rm MeV}]\quad$ &
\multicolumn{2}{c|}{$C_\sigma$ from OPE} &
\multicolumn{2}{c|}{$\vlowk$} &
\multicolumn{2}{c|}{$\vlowk$ + 2nd order} & \\ \hline\hline
\multirow{4}{*}{1.0} & \multirow{2}{*}{5} & 
2.48 & 2.26 & 1.27 & 1.24 & 0.241 & 0.241 &
$\: nn \leftrightarrow nn \nu \overline{\nu} \:$ \\
&& 3.46 & 2.81 & 1.94 & 1.76 & 0.401 & 0.394 &
$\: \nu nn \leftrightarrow \nu nn \:$ \\
& \multirow{2}{*}{10} & 
2.81 & 2.10 & 1.58 & 1.44 & 0.308 & 0.307 &
$\: nn \leftrightarrow nn \nu \overline{\nu} \:$ \\
&& 3.41 & 2.24 & 2.20 & 1.79 & 0.502 & 0.485 &
$\: \nu nn \leftrightarrow \nu nn \:$ \\ \hline
\multirow{4}{*}{1.7} & \multirow{2}{*}{5} & 
3.85 & 3.48 & 1.53 & 1.50 & 0.949 & 0.943 &
$\: nn \leftrightarrow nn \nu \overline{\nu} \:$ \\
&& 5.33 & 4.30 & 2.38 & 2.20 & 1.53 & 1.46 &
$\: \nu nn \leftrightarrow \nu nn \:$ \\
& \multirow{2}{*}{10} & 
4.32 & 3.18 & 1.91 & 1.80 & 1.21 & 1.18 &
$\: nn \leftrightarrow nn \nu \overline{\nu} \:$ \\
&& 5.21 & 3.37 & 2.76 & 2.35 & 1.84 & 1.67 &
$\: \nu nn \leftrightarrow \nu nn \:$ \\ \hline\hline
\end{tabular}
\caption{Rate of energy transfer $\Delta Q/\Delta T$
due to neutrino-pair bremsstrahlung
and absorption, $nn \leftrightarrow nn \nu 
\overline{\nu}$, of Eq.~(\ref{tpair}) and due to 
inelastic scattering, $\nu nn 
\leftrightarrow \nu nn$, of Eq.~(\ref{tscatt})
for characteristic temperatures and Fermi momenta.
Results are given without and with mean-field effects, $G_0 = 0$ and 
$G_0 = 0.8$ respectively, and for different spin relaxation rates 
$1/\tau_\sigma$ based on Fig.~\ref{fig:spin}. The 
rates are in units of $10^{33} \, {\rm erg} \, {\rm cm}^{-3} \, {\rm s}^{-1} 
\, {\rm MeV}^{-1}$ for $T=5 \mev$ and $10^{35} \, {\rm erg} \, 
{\rm cm}^{-3} \, {\rm s}^{-1} \, {\rm MeV}^{-1}$ for $T=10 \mev$.
\label{tab:etrans}}
\end{table}

Finally, we consider the rate of energy transfer $\Delta Q/\Delta T$
from neutron matter at temperature $T$ to a neutrino fluid at
temperature $T_\nu$, with $\Delta T = T-T_\nu$ and $|\Delta T| \ll T$.
The energy transfer due to neutrino-pair bremsstrahlung and 
absorption, $nn \leftrightarrow nn \nu \overline{\nu}$, is given by
\be
\frac{\Delta Q}{\Delta T} = \frac{C_{\rm A}^2
G_{\rm F}^2}{20 \pi^3} \, \frac{n}{T^2} \int\limits_0^\infty d\omega
\, \omega^7 \, e^{-\omega/T} \, S_\sigma(\omega) \,,
\label{tpair}
\ee
and for inelastic scattering, $\nu nn \leftrightarrow \nu nn$, one has
\be
\frac{\Delta Q}{\Delta T} = \frac{30 \, C_{\rm A}^2
G_{\rm F}^2 \, n \, T^3}{10 \pi^3} \, \int\limits_0^\infty d\omega
\, \omega^2 \, \bigl(12 + 6\omega/T + (\omega/T)^2\bigr) 
\, e^{-\omega/T} \, S_\sigma(\omega) \,.
\label{tscatt}
\ee
Our rates for the energy transfer are shown in Table~\ref{tab:etrans}
for characteristic temperatures, Fermi momenta and the various
cases for the structure factor. The pattern of these rates is similar
to what we found for other rates in Tables~\ref{tab:imfp} and~\ref{tab:eloss}.
In addition, for all cases we find that the energy transfer due 
to inelastic scattering is less than a factor $2$ larger compared 
to the contributions from neutrino-pair bremsstrahlung and absorption.
In contrast, Hannestad and Raffelt estimated this ratio to be
$10$~\cite{Raffelt3}. However, in making this estimate they used
Eqs.~(\ref{tpair}) and~(\ref{tscatt}) with
${\rm Im} \chi_\sigma(\om, q \to 0) \sim 1/\omega^2$,
while we find ${\rm Im} 
\chi_\sigma(\om, q \to 0) \sim 1/\om$.

\section{Concluding remarks}
\label{concl}

We have developed a unified treatment for neutrino processes in 
nucleon matter based on Landau's theory of Fermi liquids that
includes one- and two-particle-hole pair states consistently.
The contributions from two-particle-hole pair states are crucial
for neutrino-pair bremsstrahlung and absorption, for inelastic scattering,
modified Urca reactions, and axion emission. In supernovae, 
neutrino-pair bremsstrahlung and absorption dominate the 
neutrino-number changing reactions and are key to the production
of muon and tau neutrinos

Neutrino rates involving two nucleons can be 
calculated in terms of the collision integral in the Landau 
transport equation for quasiparticles. Using a relaxation time
approximation, we have solved the transport equation for density
and spin-density fluctuations and derived a general form for the
response functions. The solution includes
multiple-scattering effects and effects due to
non-zero wavelengths and recoil of the nucleons. We have applied
our approach to neutral-current processes in neutron matter, but
the generalization to isospin is straightforward. Our results for
the spin response are summarized by Eqs.~(\ref{structspin}), 
(\ref{Xsigma}), (\ref{imchi}), (\ref{imchi0}), (\ref{tauavC})
and the values of $C_\sigma$ of Fig.~\ref{fig:spin}.

We have calculated the relaxation times based on the OPE model 
and for a general representation of the quasiparticle 
scattering amplitude. For OPE, the spin relaxation rate is comparable
to the quasiparticle relaxation rate, $\tau/\tau_\sigma
= 4/3$. This highlights the importance of 
noncentral contributions to nuclear interactions. We therefore
performed more systematic calculations of these rates. In
addition, for $|\om| \tau_\sigma \gg 1$ and in the long-wavelength
limit, our result for the dynamical structure factor agrees with 
Raffelt {\it et al.}~\cite{Raffelt1,Raffelt2}.

Beyond OPE, we have calculated the relaxation times based on 
low-momentum interactions $\vlowk$ and
including second-order many-body contributions. The effects
of three-nucleon interactions are generally weaker in neutron
matter~\cite{neutmatt}, but need to be included in future work.
The OPE model significantly overestimates the strength of noncentral 
contributions, compared to low-momentum interactions $\vlowk$, for
all considered densities. Beyond the $\vlowk$ results, we have
found that second-order many-body contributions reduce the spin
relaxation rate especially at lower densities. This provides a
range in Figs.~\ref{fig:spin} and~\ref{fig:dens} for the effects
due to many-body correlations. By using spin relaxation times 
that incorporate both ``in-scattering'' and ``out-scattering''
terms in the transport equation, effects corresponding to vertex
corrections in the microscopic theory are automatically taken 
into account.

Using the spin response in the long-wavelength limit, but without
further approximations or Ansaetze for the structure factor, we 
have estimated the significance of the improved rates for neutrino 
mean free paths, energy loss and energy transfer. We have found
a reduction of these rates using modern nuclear forces compared
to OPE and consequently weak mean-field effects. In addition, for 
all cases we find that the energy transfer due to inelastic scattering
is not significantly larger than that due to neutrino-pair bremsstrahlung 
and absorption.

One may ask how good the relaxation time approximation is. Our 
choice of spin relaxation time is designed to agree with microscopic
theory in the collisionless limit, $|\omega| \tau_\sigma \gg 1$,
and at long wavelengths. For the hydrodynamic limit, $|\omega|
\tau_\sigma \ll 1$, and long wavelengths, exact solutions of 
the transport equation have been obtained, and one 
finds~\cite{exact1,exact2,BaymPethick}
\be
\frac{{\tau_\sigma}|_{\rm hydro}}{\tau} = \frac{4}{3}
\sum_{\nu=1,3,5,\ldots} \frac{2\nu+1}{\nu(\nu+1)[\nu(\nu+1)-2+2 \,
\tau/\tau_\sigma]} \,. 
\ee
The relaxation time for the hydrodynamic limit is always greater 
than or equal to that for the collisionless limit. For $\tau_\sigma/
\tau \gg 1$, ${\tau_\sigma}|_{\rm hydro}=\tau_\sigma$, while for 
$\tau_\sigma/\tau = 1$, ${\tau_\sigma}|_{\rm hydro} = (\pi^2/9) \,
\tau_\sigma$. Since for realistic nuclear interactions, $\tau_\sigma$
is significantly larger than $\tau$, this indicates that differences
between spin relaxation times in the collisionless and hydrodynamic 
limits are expected to be of order a few per cent. Consequently, 
uncertainties due to the use of the relaxation time approximation
are small compared with other uncertainties in the calculation.

The use of the quasiparticle transport equation with a collision 
term allows us to include some two-particle-hole pair states, but not 
all. Among contributions not included are terms that correspond 
to the incoherent parts of the propagator for a particle-hole pair,
that is to contributions that do not correspond to an intermediate
state containing a well-defined quasiparticle together with a 
well-defined quasihole. Moreover, there are intrinsic two-body 
contributions to hadronic weak currents. Further work is needed
to determine how important these additional contributions are.

There are numerous directions for future work. One is to explore 
mixtures of neutrons and protons. A second is to extend the 
calculations to situations when matter is less degenerate. As one
sees from our results, there is significant uncertainty in the 
effects of the medium on quasiparticle scattering amplitudes, 
since there are sizable differences between rates obtained 
with $\vlowk$ and those that include many-body contributions
to second order, and an important task is to reduce these 
uncertainties.

\begin{acknowledgments}
We thank Sonia Bacca, Katy Hally and Georg Raffelt for useful discussions,
and ECT*, the Niels Bohr International Academy, NORDITA and TRIUMF for hosting 
visits at various stages of this collaboration.
This work was supported in part by the Natural Sciences and Engineering
Research Council of Canada (NSERC). TRIUMF receives federal funding
via a contribution agreement through the National Research Council
of Canada.
\end{acknowledgments}

\end{document}